\documentclass[12pt]{article}
\usepackage{amsmath,amssymb}
\newcommand{\be}{\begin{equation}}
\newcommand{\ee}{\end{equation}}
\newcommand{\bea}{\begin{eqnarray}}
\newcommand{\eea}{\end{eqnarray}}

\def\bi {\begin{itemize}}
\def\ei {\end{itemize}}

\begin{document}
\begin{flushright}
\end{flushright}
\begin{center}
{\Large\bf Deformed Reissner--Nordstrom solutions in noncommutative gravity}

\vskip .7cm {\bf{{Pradip Mukherjee$^{a}$\footnote{Also Visiting Associate, S. N. Bose National Centre for Basic Sciences, JD Block, Sector III, Salt Lake City, Calcutta -700 098, India and\\ IUCAA, Post Bag 4, Pune University Campus, Ganeshkhind, Pune 411 007,India} and Anirban Saha$^{b}$\footnote{Also Visiting Associate, IUCAA, Post Bag 4, Pune University Campus, Ganeshkhind, Pune 411 007,India}}}

{\it $^{a}$Department of Physics, Presidency College,\\86/1 College Street, Kolkata-700073, West Bengal, India.\\
e-mail: pradip@iucaa.ernet.in; \\
$^{b}$Department of Physics, Sovarani Memorial College, Jagatballavpur,\\ Howrah - 711 408, West Bengal, India. \\
e-mail: anirban@iucaa.ernet.in .
}}
\end{center}
\vskip1cm
\begin{abstract}

 The leading order corrections to Reissner--Nordstrom solutions of the Einstein's equations on noncommutative space time have been worked out basing on a noncommutative gauge theory of gravity. From the corrcted metric the horizons have been derived and the curvature scalar is also computed.
The introduction of noncommutativity leads to the removal of the coordinate singularities. 
\end{abstract}

\vskip1cm

\section{Introduction}
\setcounter{equation}{0}There is a broad consensus that general relativity and ordinary differential geometry should be replaced by noncommutative (NC) geometry at some point between currently accessible energies of about 1 -- 10 Tev (~projected for the LHC at CERN ) and the Planck scale, which is $10^{15}$ times higher. Tremendus effort has thus been spent to understand theories defined over the NC spacetime in the recent years \cite{szabo1}. 
Significant portion of these works are devoted to construct a consistent theory of gravity on NC spacetime \cite {szabo2}.  The principal obstacle in this formulation comes from negotiating general coordinate invariance. Different approaches to the problem can be broadly classified on the manner in which the diffeomorphism invariance of general relativity has been treated in the NC setting. In \cite{5} a deformation of Einstein's gravity was studied using a construction based on gauging the noncommutative $SO(4,1)$ de sitter group and the Seiberg--Witten (SW) map \cite{SW} with subsequent contraction to $ISO(3,1)$. Construction of a noncommutative gravitational theory was proposed based on a twisted diffeomorphism algebra \cite{Aschierie,kur}. On the other hand the theory has been formulated basing on true physical symmetries \cite{CK1,MS,BMS} by resorting to a class of restricted coordinate transformations that preserve the NC algebra. The restriction corresponds to the formulation of NC gravity in the context of unimodular gravity \cite{Uni}. Remarkably, the formulations of the NC gravity theories are based mainly on the NC extensions of the gauge theories of gravity \cite{hehl}.

    Though there are yet unsolved conceptual problems \cite{12,13, CTZZ}, the development of NC gravity has reached a mature stage. For instance it has been an established fact that the leading order corrections to physically relevant quantities start from the second order in the NC parameter $\Theta$ \cite{MS,BMS}. Keeping in view the estimated order of magnitude of $\Theta$ \cite{Ani} the NC correction to gravity is indeed small. It will nonetheless be important in situations where singularities appear, as for example in black hole physics. This explains the recent interest in obtaining the corrected form of solutions of Einstein's equations in NC setting. Recently the corrections to the exterior Schwarzschild solution have been computed \cite{chichian} following \cite{5}. Since this approach gives us the first direct method of introducing such NC corrections, applications of the same to other solutions of the Einstein's equations will obviously be wellcome.

It is well-known that the Schwarzschild solution corresponds to the gravitational field of a spherically symmetric neutral mass distribution. If the distribution also carries charge the corresponding solutions are given by the Reissner--Nordstrom (R--N) metric. A natural step forward will be to compute the NC corrections to the R--N solutions. This is all the more important because it allows us to check the consistency of the formalism in the sense that tuning the charge of the distribution to zero one should obtain the NC Schwarzschild from the NC R--N solutions just as in the commutative manifold. Considering the complex issues involving the formulating NC gravity, this consistency requirement is indeed non-trivial and worthy to be addressed. Apart from this, the R--N solutions are important due to their intrinsic interest \cite{old} specifically in the context of their recent applications in Hawking radiation \cite{RW, BK1, BK2}. We therefore propose to construct the NC R-
 -N solutions following \cite{5} in the present paper.

   We assume a canonical (i.e. constant) noncommutative algebra
\begin{equation}
\left[ {x^{\mu },x^{\nu }}\right] =i\,\Theta ^{\mu \nu },
\label{3.1}
\end{equation}
Functions defined over this noncommutative space-time can be represented by ordinary functions with the product rule modified by $\star$-multiplication
\begin{equation}
\hat \phi(x) \star \hat \psi(x) = \left(\hat \phi \star \hat \psi \right)(x) = e^{\frac{i}{2}
\theta^{\alpha\beta}\partial_{\alpha}\partial^{'}_{\beta}} 
  \hat \phi (x) \hat \psi(x^{'})\big{|}_{x^{'}=x.} 
\label{star}
\end{equation}
The $\star$-multiplication is associative but not commutative.
 The space-time of noncommutative theory will be taken to be of Minkowski type, endowed with spherical noncommutative coordinates. A deformation of the gravitational field is constructed by gauging the noncommutative de Sitter $SO(4,1)$ group \cite{5} and using Seiberg-Witten (SW) map
\cite{SW}. The deformed gravitational gauge potentials (tetrad fields) $\hat{e} _{\mu }^{a}\left( {x,\Theta }\right) $ are obtained by contraction of the noncommutative gauge group $SO(4,1)$ to the Poincar\'{e} (inhomogeneous Lorentz) group $ISO(3,1)$. The fields are expanded in perturbative series where the different terms of the series are obtained from the commutative solution of the metric. We find the deformed gauge fields  up to the second order in the noncommutativity parameters $\Theta ^{\mu \nu}$ \cite{5, chichian}. The correction terms require the commutative tetrad fields of the de Sitter gauge theory of gravitation over Minkowski spacetime. We found these solutions using a spherically symmetric ansatz \cite{4} and solving the corressponding Einstein equations. From the NC tetrad fields $\hat{e} _{\mu }^{a}\left( {x,\Theta }\right) $ we construct the NC Reissner--Nordstrom metric $\hat{g}_{\mu \nu }\left( {x,\Theta }\right) $. Naturally, the nontrivial correction starts from the second order. We explicitly calculate this leading NC correction term to the Reissner--Nordstrom metric. These solutions will be used to discuss some physical consequences of the theory. 

Before proceeding further let us note that the perturbative expression of the tetrad fields only involves the commutatuive tetrad field solutions. Thus electromagnetic interaction appears in our analysis at the commutative level only. Of course NC effects will appear in the dynamics of the electromagnetic field which is however not discussed here. In our calculation we follow the convention of the signature: $-, +, +, +$ and also the units $G = c = 1$.

The organisation of our paper is as follows. Section 2 is devoted to the computation of the tetrad fields of the commutative de Sitter  $SO(4,1)$ gauge theory which leads to the R--N metric. These expressions will be subsequently used in the computation of the NC corrections. In Section 3 the results for the deformed gauge potentials (tetrad fields) valid up to the second order of the expansion in $\Theta $ are reviewed. Based on these results, we calculate the  components of the deformed Reissner--Nordstrom metric following the definition of real metric of \cite{5}. Section 4 contains our concluding remarks. 

\section{Commutative tetrad field solutions}
We review first the gauge theory of the de Sitter group SO(4,1) on a
commutative 4-dimensional Minkowski space-time endowed with the
spherically symmetric metric \cite{4}:
\begin{equation}
ds^{2}= - dt^{2} + dr^{2} + r^{2}\left( {d\theta^{2}+ \sin ^{2}\theta d\varphi ^{2}}%
\right)
\label{2.1}
\end{equation}
The $SO(4,1)$ group is 10-dimensional. We can group the gauge fields $h^A_{\mu}=(1,2,...,10)$as the four tetrad fields $e^a_{\mu}, a = 0,1,2,3$ and the six antisymmetric spin connections
$\omega ^{ab}_{\mu}, a,b = 0,1,2,3$. The field strength tensor can be separated in the torsion
\begin{equation}
F_{\mu \nu }^{a}\equiv k\,T_{\mu \nu }^{a}=k\left[ {\partial _{\mu
}e_{\nu }^{a}-\partial _{\nu }e_{\mu }^{a}+\left( \omega {_{\mu
}^{ab}\,e_{\nu }^{c}-\omega _{\nu }^{ab}\,e_{\mu }^{c}}\right)
\,\eta _{bc}}\right] , \label{2.3}
\end{equation}
and 
the curvature tensor
\begin{equation}
\begin{array}{l}
F_{\mu \nu }^{ab}\equiv R_{\mu \nu }^{ab}=\partial _{\mu }\omega
_{\nu }^{ab}-\partial _{\nu }\omega _{\mu }^{ab}+\left( {\omega
_{\mu }^{ac}\omega
_{\nu }^{db}-\omega _{\nu }^{ac}\omega _{\mu }^{db}}\right) \eta _{cd} \\
\quad \quad \quad \quad \quad +k\left( {e_{\mu }^{a}e_{\nu
}^{b}-e_{\nu }^{a}e_{\mu }^{b}}\right) 
\end{array}
\label{2.4}
\end{equation}
Under the contraction $k\to 0$ the de Sitter gauge group goes to the $ISO(3,1)$ Poincare group.
The resulting theory becomes equivalent to Einstein's theory of gravity when the torsion is set to zero.By imposing the condition of null torsion $T_{\mu \nu}^{a}=0$, one can solve for $\omega _{\mu }^{ab}(x)$ in terms of $e_{\mu }^{a}(x)$. So in this framework the spin connections are not independent fields, they are determined by the tetrads.

Now, we consider a particular form of spherically symmetric gauge fields of
the $SO(4,1)$ group given by the following Ansatz \cite{4}:
\begin{eqnarray}
e_{\mu}^{0} &=& \left(A, 0, 0, 0\right); \qquad e_{\mu}^{1} = \left(0, \frac{1}{A}, 0, 0\right);\nonumber\\
e_{\mu}^{2} &=& \left(0, 0, rC, 0\right); \qquad e_{\mu}^{3} = \left(0, 0, 0, rC \sin \theta\right)
\label{2.5}
\end{eqnarray}
\begin{equation}
\begin{array}{l}
\omega _{\mu }^{01} = \left( U, 0, 0, 0\right) ,\,\omega _{\mu
}^{12}=\left( 0, 0, W, 0\right) ,\; \omega _{\mu
}^{13}=\left( 0, 0, 0, Z \sin \theta \right) , \\
\omega _{\mu }^{23}=\left( V, 0, 0, \cos \theta \right) ,\;\omega _{\mu
}^{02} = \omega _{\mu }^{03} = \left(0, 0, 0, 0\right) ,
\end{array}
\label{2.6}
\end{equation}
where $A,\,U,\,V,\,W$ and $Z$ are functions only of the
three-dimensional radius. The non-zero components of $T_{\mu \nu
}^{a}$ and $R_{\mu \nu }^{ab}$ can be obtained using GRTensor II package of Maple as
\begin{equation}
\begin{array}{l}
T_{01}^{0}=-\frac{A\,{A}^{\prime }+U}{A},\;T_{03}^{2}= - r\,CV\,\sin
\,\theta, 
\;T_{12}^{2}= C + rC^{\prime}  - \frac{W}{A}, \\
T_{02}^{3}= r\,CV,\; T_{13}^{3}=\left(C + rC^{\prime}  - \frac{Z}{A}\right)\,\sin
\,\theta ,
\end{array}
\label{2.7}
\end{equation}
and respectively
\begin{equation}
\begin{array}{l}
R_{01}^{01}= - {U}^{\prime },\;R_{01}^{23}= -{V}^{\prime},\;R_{23}^{13}=\left(Z-W \right) \,\cos \,\theta , \\
R_{02}^{02}= - U W,\,R_{02}^{13}= V\,W,\; R_{03}^{03}=-U\,Z\,\sin \,\theta,  \\
R_{03}^{12}= - V\,Z\,\sin \,\theta ,\,R_{12}^{12}= {W}^{\prime},\;R_{23}^{\,23}= - \left( 1-Z\,W\right) \sin \,\theta , \\
R_{13}^{13}={Z}^{\prime }\,\sin \,\theta
\end{array}
\label{2.8}
\end{equation}
where ${A}^{\prime },\,{U}^{\prime },\,{V}^{\prime },\,{W}^{\prime
}$and ${Z} ^{\prime }$ denote the derivatives of first order with
respect to the $r$-coordinate.

If we use (\ref{2.7}), then the condition of null-torsion $T_{\mu
\nu }^{a}=0$ gives the following constraints:s
\begin{equation}
U=-A\,{A}^{\prime },\;V=0,\;W = Z= A \left( C + r C^{\prime}\right)
\label{2.9}
\end{equation}

We consider the case of static spherically symmetric charged matter. The Riemann tensor of the model is defined by
\begin{equation}
\tilde{R}^{\alpha\beta}_{\mu\nu} = R_{\mu \nu }^{ab}e^{\alpha}_{a}e^{\beta}_{b}
\label{riemann}
\end{equation}
The Einstein equations are \cite{Weinberg}
\begin{equation}
\tilde{R}_{\mu }^{\nu}-\frac{1}{2}\,\tilde{R}\,\delta_{\mu }^{a}= - T^{\nu}_{\mu},  \label{2.10}
\end{equation}
Where $T^{\nu}_{\mu}$ is the usual electrmagnetic energy momentum tensor
\begin{equation}
T_{\mu\nu}=2 \left(F_{\mu \lambda} F_{\nu}{}^{\lambda} - \frac{1}{4}g_{\mu\nu} 
F_{\alpha \beta}F^{\alpha \beta}\right)
\end{equation}
For our model the equations (\ref{2.10}) become 
\begin{eqnarray}
- \frac{AW^{\prime}}{rC} + \frac{1 - ZW}{r^{2}C^{2}} - \frac{AZ^{\prime}}{rC} &=&  \frac{Q^{2}}{r^{4}} \nonumber \\
\frac{WU}{rCA} + \frac{1 - ZW}{r^{2}C^{2}} - \frac{UZ}{rCA} &=& \frac{Q^{2}}{r^{4}} \nonumber \\
U^{\prime} + \frac{ZU}{rCA} - \frac{AZ^{\prime}}{rC} &=& -\frac{Q^{2}}{r^{4}}\nonumber \\
U^{\prime} + \frac{WU}{rCA} - \frac{AW^{\prime}}{rC} &=& -\frac{Q^{2}}{r^{4}}\nonumber \\
\left( W - Z \right) A &=& 0
\label{18}
\end{eqnarray}
From the above equation and (\ref{2.9})( i.e. the zero torsion constraints ) we find that the function $C$ can be conveniently chosen. Taking $C = 1$ we obtain only two independent equations
\begin{eqnarray}
- \frac{2AA^{\prime}}{r} + \frac{1 - A^{2}}{r^{2}} = \frac{Q^{2}}{r^{4}} \nonumber\\
- \frac{2AA^{\prime}}{r} + U^{\prime} = - \frac{Q^{2}}{r^{4}} 
\label{20}
\end{eqnarray}
The compatibility of these two equations gives
\begin{equation}
U^{\prime} = \frac{1 - A^{2}}{r^{2}} - \frac{2 Q^{2}}{r^{4}}
\label{21}
\end{equation}
Combining this with the first of equation (\ref{2.9}) we get the following differential equation for $A$
\begin{equation}
\left(A^{2}\right)^{\prime\prime} = \frac{2 A^{2}}{r^{2}} - \frac{2}{r^{2}} - \frac{4 Q^{2}}{r^{4}}
\label{23}
\end{equation}
The solution to (\ref{23}) is
\begin{equation}
A^{2} = 1 + \frac{\alpha}{r} + \beta r^{2} + \frac{Q^{2}}{r^{2}}
\label{24}
\end{equation}
 Here $\alpha$ and $\beta$ are arbitrary constants appearing in the complementary function of (\ref{23}). Substituting in (\ref{20}) we get $\beta = 0$. Chosing $\alpha = - 2M $ we find
\begin{equation}
A^{2} = 1 - \frac{2M}{r} + \frac{Q^{2}}{r^{2}}
\label{26}
\end{equation}
The metric $g_{\mu\nu}$ is obtained in the usual way
\begin{equation}
g_{\mu\nu}= e^{a}_{\mu}e^{b}_{\nu}\eta _{ab}
\label{cmetric}
\end{equation}
We find that
\begin{equation}
ds^{2} = - \left(1 - \frac{2M}{r} + \frac{Q^{2}}{r^{2}}\right) dt^{2} + \frac{dr^{2}}{\left(1 - \frac{2M}{r} + \frac{Q^{2}}{r^{2}}\right)} + r^{2}\left( {d\theta ^{2}+\sin ^{2}\theta d\varphi ^{2}}\right)
\label{RN}
\end{equation}
which is the R--N solution. We thus see that the ansatz (\ref{2.5}, \ref{2.6}) with $A$ given by (\ref{26}) along with the zero torsion constraints (\ref{2.9}) leads to the R--N solutions of the Einstein equations in commutative space time. Thus equations (\ref{2.5}) with (\ref{26}) are our desired solutions for the commutative tetrad fields. We will use these solutions to calculate the NC corrections in the next section.

\section{Deformed Reissner--Nordstrom Metric}
     In the above we have obtained the solutions for the tetrad fields for the de Sitter gauge theory contracted to the Poincare gauge theory in the commutative space time which leads to the R--N solutions of the Einstein equations. In this section we will use these solutions to find the R--N solutions in NC gravity. As has been mentioned earlier we follow the approach of \cite{5}.
It is possible to write the deformed tetrad fields $\hat{e}_{\mu }^{a}(x,\Theta )$ up to
the second order as \cite{5}: 
\begin{equation}
\hat{e}_{\mu }^{a}\left( {x,\Theta }\right) =e_{\mu }^{a}\left(
x\right) -i\,\,\Theta ^{\nu \rho }\,e_{\mu \nu \rho }^{a}\left(
x\right) +\Theta ^{\nu \rho }\,\Theta ^{\lambda \tau }\,\,e_{\mu \nu
\rho \lambda \tau }^{a}\left( x\right) +O\left( \Theta {^{3}}\right)
,  \label{3.10}
\end{equation}
where
\begin{eqnarray}
e_{\mu \nu \rho }^{a}&=&\frac{1}{4}\left[ {\omega _{\nu
}^{a\,c}\partial _{\rho }e_{\mu }^{d}+\left( {\partial _{\rho
}\omega _{\mu }^{a\,c}+R_{\rho \mu }^{a\,c}}\right) \,e_{\nu
}^{d}}\right] \,\eta _{c\,d},  \label{3.11}\\
e_{\mu \nu \rho \lambda \tau }^{a}&=&\frac{1}{32}\left[ {2\left\{
{R_{\tau \nu },R_{\mu \rho }}\right\} ^{a\,b}\,e_{\lambda
}^{c}-\omega _{\lambda }^{a\,b}\left( {D_{\rho }\,R_{\tau \mu
}^{c\,d}+\partial _{\rho }\,R_{\tau \mu }^{c\,d}}\right) \,e_{\nu
}^{m}\,\eta _{d\,m}}\right.\cr
&-&\left\{ {\omega _{\nu },\left( {D_{\rho }R_{\tau \mu }+\partial
_{\rho }R_{\tau \mu }}\right) }\right\} ^{a\,b}\,\,e_{\lambda
}^{c}-\partial _{\tau }\left\{ {\omega _{\nu },\left( {\partial
_{\rho }\,\omega _{\mu }+R_{\rho \mu }}\right) }\right\}
^{a\,b}\,e_{\lambda }^{c}  \label{3.12}\\
&-&\omega _{\lambda }^{a\,b}\,\partial _{\tau }\left( {\omega _{\nu
}^{c\,d}\,\partial _{\rho }e_{\mu }^{m}+\left( {\partial _{\rho
}\,\omega _{\mu }^{c\,d}+R_{\rho \mu }^{c\,d}}\right) \,e_{\nu
}^{m}}\right) \,\eta _{dm}+2\,\partial _{\nu }\omega _{\lambda
}^{a\,b}\partial _{\rho }\partial _{\tau }\,e_{\mu }^{c}\cr
&-&2\,\partial _{\rho }\left( {\partial _{\tau }\,\omega _{\mu
}^{a\,b}+R_{\tau \mu }^{a\,b}}\right) \,\partial _{\nu }\,e_{\lambda
}^{c}-\left\{ {\omega _{\nu },\left( {\partial _{\rho }\omega
_{\lambda }+R_{\rho \lambda }}\right) }\right\} ^{a\,b}\partial
_{\tau }\,e_{\mu }^{c}\cr
&-&\left. {\left( {\partial _{\tau }\,\omega _{\mu }^{a\,b}+R_{\tau
\mu }^{a\,b}}\right) \,\left( {\omega _{\nu }^{c\,d}\partial _{\rho
}e_{\lambda }^{m}+\left( {\partial _{\rho }\,\omega _{\lambda
}^{c\,d}+R_{\rho \lambda }^{c\,d}}\right) \,e_{\nu }^{m}\,\eta
_{d\,m}}\right) }\right] \,\eta _{b\,c}.\nonumber
\end{eqnarray}
We define also the complex conjugate $\hat{e}_{\mu }^{a+}\left(
{x,\Theta } \right) $\ of the deformed tetrad fields given in
(\ref{3.10}) by:
\begin{equation}
\hat{e}_{\mu }^{a}{}^{+}\left( {x,\Theta }\right) =e_{\mu
}^{a}\left( x\right) +i\,\,\Theta ^{\nu \rho }\,e_{\mu \nu \rho
}^{a}\left( x\right) +\Theta ^{\nu \rho }\Theta ^{\lambda \tau
}e_{\mu \nu \rho \lambda \tau }^{a}\left( x\right) +O\left( \Theta
{^{3}}\right) .  \label{3.13}
\end{equation}
Then we can introduce a deformed metric by the formula:
\begin{equation}
\hat{g}_{\mu \nu }\left( {x,\Theta }\right) =\frac{1}{2}\,\eta
_{a\,b}\,\left( {\hat{e}_{\mu }^{a}\star \hat{e}_{\nu
}^{b}{}^{+}+\hat{e} _{\nu }^{b}\star \hat{e}_{\mu }^{a}{}^{+}}\right)
.  \label{3.14}
\end{equation}
We can see that this metric is, by definition, a real one, even if
the deformed tetrad fields $\hat{e}_{\mu }^{a}\left( {x,\Theta
}\right) $ are complex quantities.

   We are now in a position to find the NC corrections to the R--N metric for the NC space time given by (\ref{3.1}). 
We choose the coordinate system so that the parameters $\Theta ^{\mu \nu }$ are given by
\begin{equation}
\Theta^{\mu \nu }=\left(
\begin{array}{cccc}
0 & 0 & 0 & 0 \\
0 & 0& \Theta & 0 \\
0 & -\Theta & 0 & 0 \\
0 & 0 & 0 & 0
\end{array}
\right) ,\quad \left(\mu ,\nu = 0, 1, 2, 3 \right) \label{4.1}
\end{equation}
Note that in even dimension the anti-symmetric tensor $\Theta^{\mu \nu}$ can always be rotated to a skew-diagonal form \cite{szabo1}. Our form (\ref{4.1}) further assumes vanishing noncommutativity in the time-space sector, which is quite usual in the literature. Another motivation of our choice of $\Theta^{\mu \nu}$ follows from the requirement of comparision with the results of \cite{chichian}, which provids a consistency check  of the entire formalism as we have mentioned earlier.

 The non-zero components of the tetrad fields $\hat{e}_{\mu}^{a}\left( {x,\Theta }\right) $ corresponding to this NC structure can be easily worked out using GRTensor II package of Maple. Then using the definition of the metric (\ref{3.14}) we arive at the following non-zero components of the deformed metric $\hat{g}{}_{\mu \nu}$ up to the second order in $\Theta$:
\begin{eqnarray}
\hat{g}_{1\,1}\left( {x,\Theta }\right)
&=&\frac{1}{A^{2}}+\frac{1}{4}\,\frac{{ A}^{\prime \prime
}}{A}\,{\Theta }^{2}+O( {\Theta ^{4}}) ,\nonumber \\
\hat{g}_{22}\left( {x,\Theta }\right) &=&r^{2}+\frac{1}{16}\,\left(
{ A^{2}+11\,r\,A\,{A}^{\prime }+16\,r^{2}\,{A}^{\prime
}{}^{2}+12\,r^{2}A\,{A} ^{\prime \prime }}\right) \,{\Theta }^{2}+O(
{\Theta ^{4}}) , \cr
\hat{g}_{33}\left( {x,\Theta }\right) &=&r^{2}\,\sin ^{2}\,{\theta
}\cr
&+&\frac{1 }{16}\,\left[ {4\,\left( {2\,r\,A\,{A}^{\prime
}-\,r\frac{{A}^{\prime }}{A} +\,r^{2}\,A\,{A}^{\prime \prime
}+2\,r^{2}\,{A}^{\prime }{}^{2}}\right) \,\sin ^{2}\,\theta +\cos
^{2}\theta \,}\right] \,{\Theta }^{2}+O( {\Theta ^{4}})\cr
\hat{g}_{00}\left( {x,\Theta }\right) &=&-A^{2}-\frac{1}{4}\,\left(
{2\,r\,A\,{ A}^{\prime }{}^{3}+r\,A^{3}\,{A}^{\prime \prime \prime
}+A^{3}\,{A}^{\prime \prime }+2\,A^{2}\,{A}^{\prime
}{}^{2}+5\,r\,A^{2}\,{A}^{\prime }\,{A} ^{\prime \prime }}\right)
\,{\Theta }^{2}+O( {\Theta ^{4}}) ,\nonumber
\label{4.3}
\end{eqnarray}
where ${A}^{\prime },\,{A}^{\prime \prime },\,{A}^{\prime \prime
\prime }$ are first, second and third derivatives of $A(r)$, given by (\ref{26}), respectively. The same expressions will be useful in our work with only one change i.e. $A\left(r \right)$ is now given by (\ref{26}). One can now compute the corrections to the R--N metric for NC gravity. The explicit form of the non-zero components are 
\begin{eqnarray}
\hat{g}_{00}&=& -\left( {1-\frac{2M }{r} + \frac{Q^{2}}{r^{2}}}\right) - \frac{1}{r^{6}}\left[Mr^{3} - \frac{11 M^{2} + 9 Q^{2}}{4}r^{2} - \frac{17 M Q^{2}}{4} r - \frac{7 Q^{4}}{2}\right]\Theta^{2}+ O(\Theta^{4}) \nonumber \\
\hat{g}_{11}&=& \frac{1}{\left( {1-\frac{2M }{r} + \frac{Q^{2}}{r^{2}}}\right)} + \frac{\left[- 2 M r^{3} + 3\left(M^{2} + Q^{2}\right) r^{2} - 6 M Q^{2} r + 2 Q^{4} \right]}{4 r^{2} \left(r^{2} - 2 M r + Q^{2}\right)^{2}} \Theta^{2} + O(\Theta^{4})\nonumber\\
\hat{g}_{22}&=& r^{2} + \frac{1}{16}\left[1 - \frac{15 M}{r} + \frac{26 Q^{2}}{r^{2}} + \frac{4 \left(M r - Q^{2}\right)^{2}}{r^{2} \left(r^{2} - 2 M r + Q^{2}\right)}\right]\Theta ^{2} +O(\Theta ^{4}) \nonumber \\
\hat{g}_{33}&=& r^{2} \sin^{2}\theta + \frac{1}{16}\left[\frac{4 r^{2} \left(M^{2} - M r\right) + 8 Q^{2} \left( r^{2} - 2 M r\right) + 8 Q^{4} }{r^{2} \left(r^{2} - 2 M r + Q^{2}\right)^{2}} \sin^{2} \theta + \cos^{2} \theta\right]\Theta ^{2}+ O(\Theta ^{4}) \nonumber \\
\label{4.5} 
\end{eqnarray}
We thus reach our desired results. Certain observations are due at this point. 
\begin{enumerate}
\item If we substitute $Q = 0$ in our expressions (\ref{4.5}) the solutions exactly reduces to the NC Schwarzschild solutions \cite{chichian}. It is well-known that the R--N metric goes over to the Schwarzschild metric in the limit $Q \to 0$ for commutative space time. That the same correspondence prevails for NC space time as well is indeed gratifying considering the complexities involved in the construction.
\item The NC tetrad fields contain non-vanishing terms first order in the NC parameter $\theta$. 
There is however no ambiguity because the complex tetrad fields are not physical observables. It is the metric which are physically observable and our corrections to the metric (\ref{4.5}) indeed starts from the second order. This result is thus consistent with the general observation that there is no observable first order correction to NC gravity \cite{MS, CK2, deliduman, HSK}. 
\end{enumerate}
Note that attempts to incorporate NC effects for the charged black hole was made earlier in the literature \cite{early} but they assumed the commutative expression for the metric and introduced noncommutativity afterwards. It can be definitely claimed that our results provide the first rigorous expressions for NC corrections to the R--N metric. 
\section{Some physical consequences of the NC correction}
In this section we will derive some physical results steming from the NC corrections presented above. Specifically we will discuss how the NC corrections affect the R--N horizon radii. Also we will derive an expression for the NC curvature scalar upto second order in the NC parameter. This later computation is necessary to butress our assertion that the complex first order term in the NC tetrad does not contribute to the physical results of the theory.
\subsection{Horizon corrections}
In commutative space-time we can identify the event horizons by following radial null curves and locating the radious at which $\frac{dt}{dr}$ becomes infinity. Following this and remembering that our event horizons should go to the commutative results in the limit $\theta \to 0$ we define the event horizons of the NC R--N metric from $g_{00} = 0$.
From our NC R--N solutions (\ref{4.5}) it is straightforward to derive 
\begin{eqnarray}
r^{2} - 2 mr + Q^{2} = -\frac{\Theta^{2}}{4 r^{4}}\left(4 m r^{3} - 11 m^{2} r^{2} + 9 Q^{2} r^{2} - 17 m Q^{2} r + 14 Q^{4}\right)
\label{H1}
\end{eqnarray}
the solutions to which give the horizon redii. Naturally we look for solutions correct upto second order in $\Theta$. The required solutions are 
\begin{eqnarray}
r_{+} &=& M + \sqrt{M^{2} - Q^{2}} + \frac{\Theta^{2}}{2}\frac{A_{+}}{\sqrt{M^{2} - Q^{2}}} \nonumber\\
r_{-} &=& M - \sqrt{M^{2} - Q^{2}} - \frac{\Theta^{2}}{2}\frac{A_{-}}{\sqrt{M^{2} - Q^{2}}}
\label{H2}
\end{eqnarray}
with $A_{+}$ and $A_{-}$ given by 
\begin{eqnarray}
A_{+} &=& \frac{6 M^4 + 10 M^3 \left( M^2 - Q^2 \right)^{(1/2)} + 36 Q^2 M^2
 - 4 M \left( M^2 - Q^2 \right)^{(3/2)}}{4 \left(M + \sqrt{M^2 - Q^2}\right)^4}\nonumber\\
&& + \frac{35 Q^2 M \left( M^2 - Q^2 \right)^{(1/2)} + 5 Q^4}{4 \left(M + \sqrt{M^2 - Q^2}\right)^4}
 \nonumber\\
A_{-} &=& \frac{6 M^4 - 10 M^3 \left( M^2 - Q^2 \right)^{(1/2)} + 36 Q^2 M^2
 + 4 M \left( M^2 - Q^2 \right)^{(3/2)}}{4 \left(M - \sqrt{M^2 - Q^2}\right)^4}\nonumber\\
&& - \frac{35 Q^2 M \left( M^2 - Q^2 \right)^{(1/2)} + 5 Q^4}{4 \left(M - \sqrt{M^2 - Q^2}\right)^4}
\label{H3}
\end{eqnarray}
Note that these solutions properly map to the familear (commutative) R--N horizon 
\begin{eqnarray}
r_{\pm} = M \pm \sqrt{M^{2} - Q^{2}} 
\label{H4}
\end{eqnarray}
in the limit $\Theta \to 0$.
As a result of the NC effect the distance between the event horizon redii increases.

 A crucial point needs to be  mentioned before passing. From our construction it appears that at $r_{\pm}$ the determinant of the metric vanishes indicating genuine singularity. This result will indeed be paradoxical because there is no such singularity in the commutative limit. However, note that the underlying space-time is noncommutative and the determinant must be defined with respect to $\star$-multiplication and not ordinary multiplication. From the difinition of $\star$-multiplication it can be appreciated that new $\Theta$-dependent terms will come accompanying derivatives of the metric elements resulting into non-zero value of the determinant. This determinant should be calculated from the definition (\ref{3.14}) of the metric elements and using the determinant of the NC tetrads given by 
\begin{eqnarray}
\mbox{det}_{\star}(e^a_\mu(x))\stackrel{\rm{def}}{=}\frac{1}{4!}\epsilon^{\mu \nu \rho \sigma}
\epsilon_{abcd}e^a_\mu(x)\star e^b_\nu(x)\star e^c_\rho(x)\star e^d_\sigma(x)
\label{detg}
\end{eqnarray} 
We do not give the detailed expression of this determinant since it will be rather cumbersome and will not be used in the sequel.
\subsection{The NC scalar curvature}
We propose to work out the NC scalar curvature from the NC tetrad and spin connections. The later can be expanded as \cite{5}
\begin{eqnarray}
\hat{\omega}{}^{ab}{}_{\mu} = \omega{}^{ab}{}_{\mu} - i \Theta^{\nu \rho} \omega{}^{ab}{}_{\mu \nu \rho} + \Theta^{\nu \rho} \Theta^{\lambda \tau} \omega{}^{ab}{}_{\mu \nu \rho \lambda \tau}
\label{R1}
\end{eqnarray}
where $\omega^{ab}{}_{\mu \nu \rho}$ and $\omega{}^{ab}{}_{\mu \nu \rho \lambda \tau}$ are the first and second order corrections respectively which are given by 
\begin{eqnarray}
\omega^{ab}{}_{\mu \nu \rho} &=& \frac{1}{4}\,\,\left\{ {\omega _{\nu },\,\partial _{\rho }\omega_{\mu }+ R_{\rho \mu}}\right\} ^{ab} \nonumber\\
\omega{}^{ab}{}_{\mu \nu \rho \lambda \tau} &=& \frac{1}{32} \,\,\left( {-\left\{ {\omega _{\lambda },\partial
_{\tau }\left\{ {\omega _{\nu },\partial _{\rho }\omega _{\mu
}+R_{\rho \mu }}\right\} }\right\} +2\left\{ {\omega _{\lambda
},\left\{ {R_{\tau \nu },R_{\mu \rho }}\right\} } \right\} }\right. \nonumber \\
&-&\left\{ {\omega _{\lambda },\left\{ {\omega _{\nu },D_{\rho
}R_{\tau \mu }+\partial _{\rho }R_{\tau \mu }}\right\} }\right\}
-\left\{ {\left\{ { \omega _{\nu },\partial _{\rho }\omega _{\lambda
}+R_{\rho \lambda }} \right\} ,\left( {\partial _{\tau }\omega _{\mu
}+R_{\tau \mu }}\right) } \right\} \cr &+&\left. {2\left[ {\partial
_{\nu }\omega _{\lambda },\partial _{\rho }\left( {\partial _{\tau
}\omega _{\mu }+R_{\tau \mu }}\right) }\right] \,}\right)
^{ab},
\label{R2}
\end{eqnarray}
Similarly the NC Riemann tensor is also expanded as 
\begin{equation}
\hat{R}^{ab}_{\mu\nu}=R^{ab}_{\mu\nu}+i\Theta^{\rho\tau}R^{ab}_{\mu\nu\rho\tau}+\Theta^{\rho\tau}\Theta^{\kappa\sigma}
R^{ab}_{\mu\nu\rho\tau\kappa\sigma}+O(\Theta^{3})\,,
\label{R3}
\end{equation}
where
\begin{equation}
R^{ab}_{\mu\nu\rho\tau}=\partial_{\mu}\omega^{ab}_{\nu\rho\tau}+(\omega^{ac}_{\mu}\omega^{db}_{\nu\rho\tau}+
\omega^{ac}_{\mu\rho\tau}+\omega^{db}_{\nu}-\frac{1}{2}\partial_{\rho}\omega_{\mu}^{ac}\partial_{\tau}\omega_{\nu}^{db})
\eta_{cd}-(\mu\leftrightarrow \nu)
\label{R4}
\end{equation}
and
\begin{equation}
R^{ab}_{\mu\nu\rho\tau\kappa\sigma}=\partial_{\mu}\omega^{ab}_{\nu\rho\tau\kappa\sigma}+
(\omega^{ac}_{\mu}\omega^{db}_{\nu\rho\tau\kappa\sigma}+
\omega^{ac}_{\mu\rho\tau\kappa\sigma}+\omega^{db}_{\nu}-\omega^{ac}_{\mu\rho\tau}\omega^{db}_{\nu\kappa\sigma}-
\frac{1}{4}\partial_{\rho}\partial_{\kappa}\omega^{ac}_{\mu}\partial_{\tau}\partial_{\sigma}\omega_{\nu}^{db})\eta_{cd}-(\mu\leftrightarrow
\nu)
\label{R5}
\end{equation}
The NC scalar curvature will be obtained as 
\begin{equation}
\hat{R}=\hat{e}^{\mu}_{ a}\star \hat{R}^{ab}_{\mu\nu}\star
\hat{e}^{\nu}_{b}
\label{R6}
\end{equation}
where $\hat{e}^{\mu}_{a}$  is the $\star$-inverse of $\hat{e}^{a}_{\mu}$, i.e. 
\begin{eqnarray}
\hat{e}^{\mu}_{a}\star\hat{e}^{b}_{\mu}=\delta^b_a
\label{R0}
\end{eqnarray}
 The general expression of the scalar curvature, expanded in powers of $\Theta$, is
\begin{eqnarray}
\hat{R}=R &+& \Theta^{\rho\tau}\Theta^{\kappa\sigma}(e^{\mu}_{a}R^{ab}_{\mu\nu\rho\tau\kappa\sigma}e^{\nu}_{b}+
e^{\mu}_{a\rho\tau\kappa\sigma}R^{ab}_{\mu\nu}e^{\nu}_{b}+e^{\mu}_{a}R^{ab}_{\mu\nu}e^{\nu}_{b\rho\tau\kappa\sigma}
-e^{\mu}_{a\rho\tau}R^{ab}_{\mu\nu}e^{\nu}_{b\kappa\sigma}\cr
&-&e^{\mu}_{a\rho\tau}R^{ab}_{\mu\nu\kappa\sigma}e^{\nu}_{b}
-e^{\mu}_{a}R^{ab}_{\mu\nu\rho\tau}e^{\nu}_{b\kappa\sigma})+O(\Theta^{4})
\label{R7}
\end{eqnarray}

In the above expression the quantities $\hat{e}^{\mu}_{a}$ are yet to be determind. We write 
\begin{equation}
\hat{e}^{\mu}_{
a}=e^{\mu}_{a}-i\Theta^{\nu\rho}e^{\mu}_{a\nu\rho}+\Theta^{\nu\rho}\Theta^{\kappa\sigma}e^{\mu}_{a\nu\rho\kappa\sigma}
+O(\Theta^{3})
\label{R8}
\end{equation}
where the corrections $e^{\mu}_{a\rho\tau}$ and $e^{\mu}_{a\rho\tau\kappa\sigma}$ are obtained 
using the defining equation (\ref{R0}) as
\begin{eqnarray}
e^{\mu}_{a\nu\rho} & = & - e^{\alpha}{}_{a} e^{b}{}_{\alpha \nu \rho} e^{\mu}{}_{b} + \frac{1}{2}\partial_{\nu} e^{\alpha}{}_{a}\partial_{\rho} e^{b}{}_{\alpha} e^{\mu}{}_{b} \nonumber\\
e^{\mu}_{a\nu \rho \lambda \tau} & = & - e^{\alpha}{}_{a} e^{b}{}_{\alpha \nu \rho \lambda \tau} e^{\mu}{}_{b} + e^{\alpha}{}_{a \nu \rho} e^{b}{}_{\alpha \lambda \tau} e^{\mu}{}_{b} \nonumber\\&+& 
\frac{1}{4}\partial_{\nu}\partial_{\lambda} e^{\alpha}{}_{a}\partial_{\rho}\partial_{\tau}e^{b}{}_{\alpha} e^{\mu}{}_{b} - \frac{1}{2}\left(\partial_{\nu}e^{\alpha}{}_{a}\partial_{\rho}e^{b}{}_{\alpha \lambda \tau} e^{\mu}{}_{b} +  \partial_{\nu}e^{\alpha}{}_{a \lambda \tau}\partial_{\rho}e^{b}{}_{\alpha} e^{\mu}{}_{b}\right)
\label{R9}
\end{eqnarray}
The noncommutative scalar curvature for the Reissner-Nordstr\"{o}m
de Sitter solution is then obtained in the form:
\begin{eqnarray}
\hat{R} &=& - \frac{\Theta^{2}}{16\left\{r^8 \sin^2 \theta \left(r^2 - 2 M r + Q^2\right)\right\}} \times\nonumber\\
&& \left[4Q^4 - 36Mr^3 Q^2 - 8 M r Q^2 + 18 Q^4 r^2 + 4 r^2 Q^2 + 18 r^4 Q^2 \right. \nonumber\\
&&+ \left.  64 r^5 M \sin^2 \theta - 106 r^4 Q^2 \sin^2 \theta - 260 M^2 r^4 \sin^2 \theta + 256  M^3 r^3 \sin^2 \theta \right. \nonumber\\
&&-\left. 105 Q^4 r^2 \sin^2 \theta + 29 Q^6 \sin^2 \theta + 455 M r^3 Q^2 \sin^2 \theta \right.
\nonumber\\
&&-\left.  458 M^2 r^2 Q^2 \sin^2 \theta + 125 M r Q^4 \sin^2 \theta \right]
\label{R10}
\end{eqnarray}
We can draw the following infarences from the above expression for the NC curvature sclar $\hat{R}$:
\begin{enumerate}
\item The non-zero value is entirely an NC effect. This is consistent with  the fact that the curvature scalar vanishes in the commutative model. Furthermore the leading order NC correction is second order in $\Theta$, same as we have found for the NC corrections to the metric. 
\item As a further consistency check we note that for $Q = M = 0$ the NC scalar curvature vanishes at least to the order calculated here. Since this limit corresponds to empty space-time the curvature is expected to vanish.  
\item At the corrected horizon (\ref{H2}) the NC scalar curvature is well behaved since the denominator does not vanish there. This can be seen by substituting (\ref{H2}) in (\ref{R10}). So it appears that introduction of  noncommutativity removes the coordinate singularities in R--N solutions. However, it should be remembered  that we arrive at this conclusion in a perturbative framework. 
\end{enumerate}
\section{Conclusion}
We have calculated the leading order noncommutative (NC) corrections to the Reissner--Nordstrom (R--N) solutions based on the formulation of NC gravity in \cite{5}. The solutions to the tetrad fields for a static spherically symmetric charged matter distribution in the NC space time have been worked out. These solutions have been used to find the NC tetrads. Here the NC tetrad is expressed in a commutative equivalent approach by treating the NC gravity theory as a contraction of NC deSitter gauge theory using the celebrated Seiberg--Witten (SW) map technique. The solutions come as a power series in the NC parameter. We retain terms upto second order with the hindsight that any non-trivial physical correction to NC gravity starts from the second order in the NC parameter \cite{MS, BMS}. Using these solutions we construct the NC R--N metric. Note that these are the first occurence of such solutions which are based on a rigorous formulation of NC gravity. The leading order corre
 ction is found to be in the second order as expected. By tuning the charge parameter we get back the Schwarzschild solutions \cite{chichian} from our results. This demonstrates that the correspondence between the Schwarzschild and the R--N solutions, so well-known in the commutative perspective, holds also for the NC space time. 

From the expressions of the corrected metric we have obtained the corrected forms of the R--N horizons. Starting from the NC solutions of the tetrad and spin connections we have derived the expression of curvature scalar. The coordinate singularities are seen to be removed due to the introduction of noncommutativity. From the corrected forms of the R--N metric and the  corresponding scalar curvatur it is evident that the spherical symmetry is broken which is expected due to the introduction of constant noncommutativity in the $r-\theta$ direction. It will be very useful if one can restore this symmetry in the NC framework. Also a detailed study of the singularities will be welcome.

\vskip 0.5cm {\bf{Acknowledgements}}
The authors like to acknowledge the excellent hospitality of IUCAA where the work was done.

\end{document}